\documentclass[pdflatex,sn-mathphys]{sn-jnl}
\usepackage[T1]{fontenc}
\usepackage{graphicx}%
\usepackage{multirow}%
\usepackage{amsmath,amssymb,amsfonts}%
\usepackage{amsthm}%
\usepackage{mathrsfs}%
\usepackage[title]{appendix}%
\usepackage{xcolor}%
\usepackage{textcomp}%
\usepackage{manyfoot}%
\usepackage{booktabs}%
\usepackage{algorithm}%
\usepackage{algorithmicx}%
\usepackage{algpseudocode}%
\usepackage{listings}%
\usepackage{lipsum}         % Lorem Ipsum
\usepackage{mathtools}
\usepackage{tabularx}
\usepackage{bm}
\usepackage{float}
\usepackage{siunitx} % UNITS
\normalbaroutside
\usepackage{tikz}
\usetikzlibrary{decorations.pathreplacing}
\definecolor{darkslategray}{RGB}{47,79,79}
\definecolor{lightslategray}{RGB}{119,136,153}
\definecolor{darkyellow}{RGB}{102,102,0}
\definecolor{da rkgreen}{RGB}{0,100,0}
\usepackage{pgfplots}
\usepackage{overpic}
\usetikzlibrary{patterns,arrows.meta,shapes.geometric, positioning, fit, calc, decorations.pathreplacing} 
\pgfplotsset{compat=1.16}
\newtheorem{theorem}{Theorem}
\begin{document}

%%=============%%
%% title %%
%%=============%%

\title{A Fourier Neural Operator Approach for Modelling Exciton-Polariton Condensate Systems}

%%=============%%
%% authors %%
%%=============%%

\author*[1]{\fnm{Yuan}~\sur{Wang}}\email{yw2e17@soton.ac.uk}
\equalcont{These authors contributed equally to this work.}

\author[2]{\fnm{Surya}~\sur{T.~Sathujoda}}
\equalcont{These authors contributed equally to this work.}

\author[1]{\fnm{Krzysztof}~\sur{Sawicki}}

\author[1]{\fnm{Kanishk}~\sur{Gandhi}}

\author[3]{\fnm{Angelica}~\sur{I~Aviles-Rivero}}

\author[4]{\fnm{Pavlos}~\sur{G.~Lagoudakis}}

\affil[1]{\orgdiv{School of Physics and Astronomy}, \orgname{University of Southampton}, \orgaddress{\city{Southampton}, \postcode{SO17 1BJ}, \country{United Kingdom}}}

\affil[2]{\orgdiv{Department of Physics}, \orgname{University of Cambridge}, \orgaddress{\city{Cambridge}, \postcode{CB3 0HE}, \country{United Kingdom}}}

\affil[3]{\orgdiv{Yau Mathematical Sciences Center}, \orgname{Tsinghua University}, \orgaddress{\city{Beijing}, \postcode{100084}, \country{China}}}

\affil[4]{\orgdiv{Hybrid Photonics Laboratory}, \orgname{Skolkovo Institute of Science and Technology, Territory of Innovation Center Skolkovo}, \orgaddress{\city{Bolshoy Boulevard 30, building 1, Moscow}, \postcode{121205}, \country{Russia}}}

%%=============%%
%% abstract %%
%%=============%%

\abstract{A plethora of next-generation all-optical devices based on exciton-polaritons have been proposed in latest years, including prototypes of transistors, switches, analogue quantum simulators and others. However, for such systems consisting of multiple polariton condensates, it is still challenging to predict their properties in a fast and accurate manner. The condensate physics is conventionally described by Gross-Pitaevskii equations (GPEs). While GPU-based solvers currently exist, we propose a significantly more efficient machine-learning-based Fourier neural operator approach to find the solution to the GPE coupled with exciton rate equations, trained on both numerical and experimental datasets. The proposed method predicts solutions almost three orders of magnitude faster than CUDA-based solvers  in numerical studies, maintaining the high degree of accuracy. Our method not only accelerates simulations but also opens the door to faster, more scalable designs for all-optical chips and devices, offering profound implications for quantum computing, neuromorphic systems, and beyond.}

\maketitle

%%=============%%
%% introduction %%
%%=============%%
\section*{Introduction}\label{sec1}
Over the decades a wide range of all‐optical devices, from switches~\cite{Amo2010, Giorgi2012,Gao2012, Dreismann2016, Ma2020, Feng2021, Chen2022} and transistors~\cite{Ballarini2013, Zasedatelev2019, Zasedatelev2021} to analogue quantum simulators~\cite{Berloff2017, Lagoudakis2017} and neuromorphic computing~\cite{Opala2019, Ballarini2020, Mirek2021, Ghosh2021, opala2022training, opala2023harnessing}, have been reported. In particular, exciton‐polariton–based devices have emerged, capitalizing on the nonlinearities and unique propagation properties of these quasiparticles~\cite{Kasprzak2006}. A notable example is the optically activated transistor switch. Initially designed for cryogenic conditions using polariton condensates~\cite{Ballarini2013}, recent advances now enable ambient operation~\cite{Zasedatelev2019, Zasedatelev2021}. Further progress of all-optical devices necessitates the development of precise and adaptable simulation tools. Just as Electronic Design Automation (EDA) played a pivotal role in the evolution of chip design, there is a pressing need for emulators that can capture the rich nonlinear characteristics inherent in optical devices. However, accurately predicting the behavior of systems with multiple coupled polariton condensates such as polariton chains, lattices, or graphs is a challenging task, and its complexity grows dramatically with the number of condensates in the system. In this connection, rapid development of machine learning techniques holds great potential for solving non-trivial many-body problems and offers a chance to revolutionise polaritonics, along with many other fields.

\begin{figure*}[htbp]
\centering
\begin{minipage}{0.2\textwidth} 
  \hspace*{-1.5cm}
  \begin{overpic}[scale=1.0,unit=1cm]{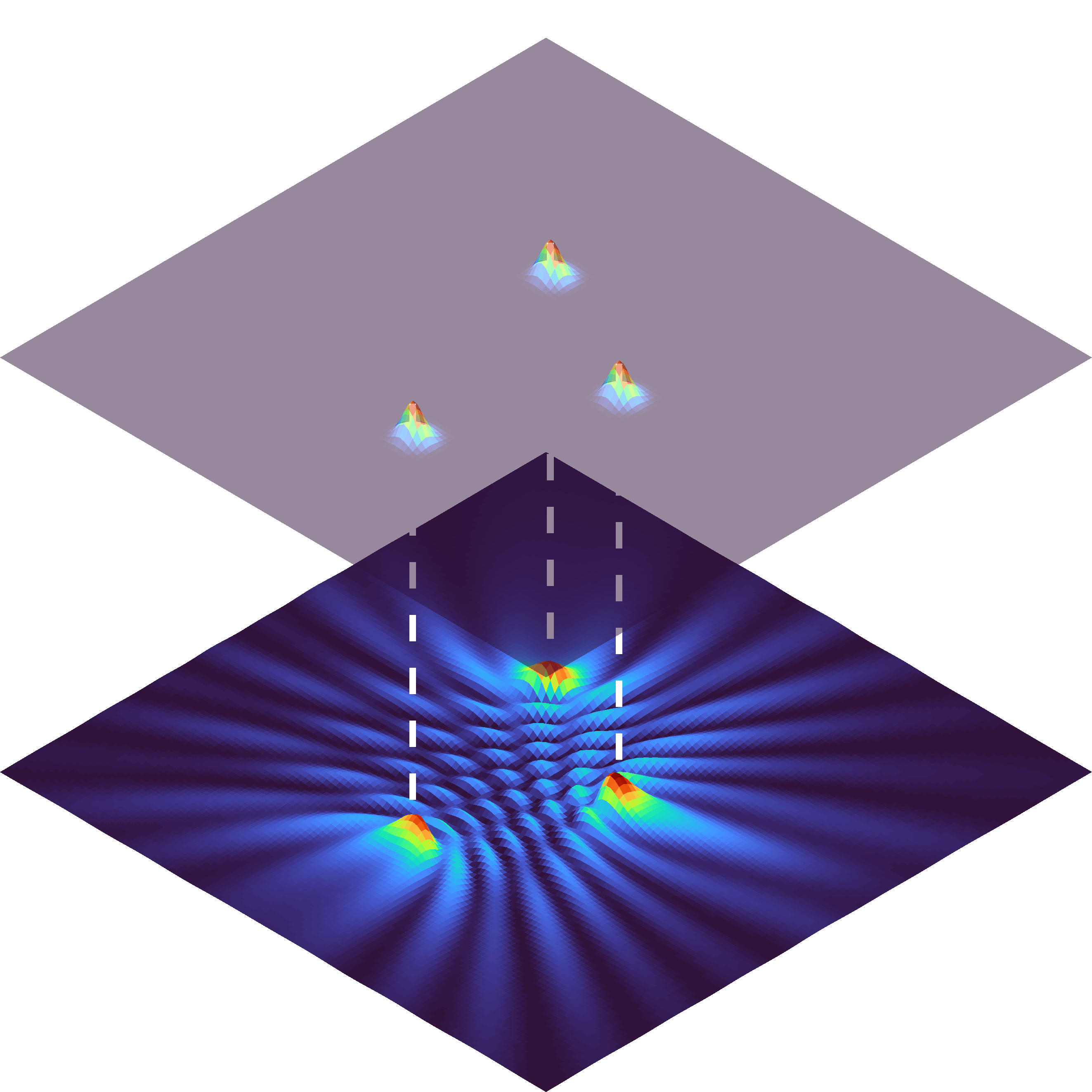}
  % Index a and b
      \put(0, 100){\color{black} \textbf{a}}
        \put(90, 100){\color{black} \textbf{b}}
    \put(76.5, 85){\color{black}\footnotesize Nonresonant pump}
    \put(75.5, 86.5){\color{black}\vector(-1,0){6}} 
    \put(119, 86.5){\color{black}\vector(1,0){6}} 
    \put(82.5,12.7){\color{black}\footnotesize Condensate density}
    \put(82.0, 14.2){\color{black}\vector(-1,0){5.8}}
    \put(126.0, 14.2){\color{black}\vector(1,0){5.8}}
    \put(85, 54){\color{black}\footnotesize Lower polariton branch}
    \put(111, 52.5){\color{black}\vector(0,-1){7.5}}
    \put(176, 21.0){\color{black}\footnotesize Reservoir}
    \put(185, 25.5){\color{black}\vector(-1,1){13}}
  \end{overpic}
\end{minipage}
\begin{minipage}{0.5\textwidth}
\hspace*{0.5cm}
\begin{tikzpicture}[
    declare function={
        hbarmevps=0.658212000000000;
        mC=0.284281505180420;
        mE=3.411378062165041e+03;
        OmegaR=6.077069394055410;
        minusdetunning=-4.0;
        EnergyC(\x)=minusdetunning + hbarmevps^2*\x^2 / (2.0*mC);
        EnergyE(\x)=hbarmevps^2*\x^2 / (2.0*mE);
        Delta(\x)=(EnergyC(\x) - EnergyE(\x)) / hbarmevps;
        OmegaL(\x)=0.5*(EnergyC(\x) + EnergyE(\x)) - hbarmevps*sqrt(0.25*Delta(\x)^2 + OmegaR^2);
    }
]
    \begin{axis}[
        xlabel=\footnotesize{Wavevector},
        ylabel=\footnotesize{Energy},
        axis y line=middle, % y-axis in the center
        axis x line=bottom,
        enlargelimits=true,
        xtick=\empty, % Remove x ticks
        ytick=\empty, % Remove y ticks
        xmin=-10, xmax=10,
        ymin=-7, % Adjust if needed
        ymax=10,  % Adjust if needed
        x axis line style={at={(axis description cs:0,-0.1)}}, % moves x-axis down
    ]
    \addplot[black, very thick, domain=-10:10, samples=100] {OmegaL(\x)};
    \draw[fill=gray, opacity=0.5] (axis cs:0,8.0) ellipse [x radius=3, y radius=0.7]; % Top ellipse
    \draw[fill=gray, opacity=0.5] (axis cs:0,-6.5) ellipse [x radius=1.75, y radius=0.5]; % Bottom ellipse
    \draw[fill=gray, opacity=0.5,rotate around={30:(axis cs:7,7)}] (axis cs:1.2,0.95) ellipse [x radius=1.5, y radius=0.5]; % Rotated ellipse on the right side

    % Dashed Arrows
    \draw[-{Latex[length=2mm]}, thick, dashed] (axis cs:3.0,7.5) to[bend left=35] (axis cs:6.0,0.0);
    \draw[-{Latex[length=2mm]}, thick, dashed] (axis cs:5.5,-2.0) to[bend left=35] (axis cs:2.0,-6.5);
    \end{axis}
\end{tikzpicture}
\end{minipage}
\caption{\label{illustration_condensates} \textbf{Comparison of pump profiles and wavefunction density with scattering process illustration.} \textbf{a} The upper layer shows the nonresonant pump profile featuring three Gaussian spots, while the lower one shows the wavefunction density of the condensates at the final time. Three white dashed lines indicate the central positions of the pump regions and align with their corresponding locations on the condensate density map. \textbf{b} Depiction of the scattering process, tracing the transition from the hot electron-hole plasma phase, through the reservoir cooling phase, to the scattering in the condensates. Only the lower polariton branch of the polariton energy mode is shown here.}
\end{figure*}

The microcavity exciton-polariton (hereafter polariton) system~\cite{Weisbuch1992}, the platform in which the above-mentioned all-optical devices are realized, consists of two main strongly coupled components: excitons confined in an active material and photons trapped in a microcavity. One of the unique features of polaritons resulting from their bosonic nature and low effective mass is their ability to form condensates, i.e., a macroscopic coherent quantum state~\cite{Kasprzak2006}. It has recently been shown that, in the non-resonantly pumped case, due to ballistic propagation, two or more spatially separated polariton condensates can interact over a large distance forming many-body systems such as dyad~\cite{Tosi2012, Toepfer2020}, chain~\cite{Pickup:NatCommun2020, Dovzhenko:PRB2023}, lattice~\cite{Berloff2017, Toepfer2021} or graph~\cite{Berloff2017} (see example in Fig.~\ref{illustration_condensates}{\bf a}). In a non-resonant scheme (Fig.~\ref{illustration_condensates}{\bf b}), polariton condensation occurs when a hot electron-hole plasma cools, forming excitons along the lower polariton branch. With increasing the photonic component, interactions with phonons and exciton-exciton scattering lead to a bottleneck region. Momentum is redistributed by parametric scattering. Once the threshold is exceeded, condensates are formed, triggering ballistic propagation~\cite{Wouters:PRB2008, Wertz2012}.

Advances in semiconductor microcavity fabrication and spatial light modulators (SLMs) have enabled effective pump profile manipulation. This versatility reveals diverse nonlinear phenomena and photoluminescence outputs that can be used to engineer quantum fluids of light~\cite{Carusotto2013}. Extensive studies have been conducted on the ability of non-resonant optical techniques to manipulate the motion of condensate polaritons, such as customized momentum distributions~\cite{Assmann2012}, condensate amplifiers~\cite{Wertz2012, Niemietz2016}, waveguides~\cite{Schmutzler2015, Cristofolini2018, Wang2021, Aristov2023}, vertical coupling of condensates~\cite{Diederichs:Nature2006, Sawicki:Nanophotonics2021}, next-nearest-neighbour coupling~\cite{Dovzhenko:PRB2023}, directional superfluids near equilibrium~\cite{Barkhausen2021}, manipulation of polariton properties by external magnetic field~\cite{Sciesiek:CommunMat2020, Whittaker:PRA2021, Sawicki:PRB2024}. Moreover, the ability to manipulate the condensate flow through reservoir engineering shows its potential for quantum computing~\cite{Kavokin2022}. The coupled condensates represent large-scale coherent states, distinct from the initial nonresonant inputs. Among all systems, polariton graphs pose a significant challenge but have shown potential in solving complex optimisation problems and simulating physical models such as Ising, XY, and Heisenberg systems~\cite{Berloff2017, Kalinin:PRB2020}. The complexity, diversity and irregularity of this system require the use of efficient solvers. We address and propose a machine-learning-based Fourier Neural Operator (FNO) approach that allows for the effective prediction of the emission pattern of complex polariton systems.

To this aim, a robust solution to the Gross-Pitaevskii equation (GPE)~\cite{Wouters2007}, which is used to theoretically describe condensates, is required. While parallel computing powered and GPU-based GPE solvers tailored for both uniform~\cite{Loncar2016, Schloss2018, Wilson2019, Smith2022} and non-uniform meshes~\cite{Kivioja2022} currently exist, we aim to adopt an even more efficient machine-learning (ML) based solver that is intended to accelerate significantly the computational process, especially in the context of designing extensive polariton networks (see case with $1024\times 1024$ grid size~\cite{Wang2022}). Thus far, a wide variety of ML architectures have been proposed to approximate solutions to general classes of partial differential equations (PDEs). These range from convolution-based methods, such as variants of the U-Net architecture~\cite{UNet2015Ronneberger}, to operator-learning methods such as Deep Operator Networks (DeepONets~\cite{Lu_2021_DeepONet}), Graph Neural Operators~\cite{li2020neural}, Multipole Graph Neural Operators~\cite{li2020multipole}, Fourier Neural Operators~\cite{li2021fourier} and Physics-informed Neural Operators~\cite{Li2021}. Though convolution-based methods have shown promise regarding the accuracy of future state predictions, they fail to scale in computing efficiency to larger systems, even with recent advancements~\cite{sathujoda2023physicsinformed}. Operator-learning methods overcome this bottleneck by learning mappings between infinite-dimensional spaces, allowing them to predict solutions at different discretisation at a similar speed. 

In this work, we study the application of the FNO architecture to approximate solutions to the GPE. We are interested in this specific variant of Neural Operator as its mathematical formulation relates very closely to the Split-Step Fourier Method (SSFM), which is the numerical method used to solve the GPE in this instance. Moreover, FNOs have shown widespread success in application to many other areas of physics and engineering~\cite{gopakumar2023fourier, WEN2022104180, 10.1190/geo2022-0268.1, LI2022100389}. We also use the high-quality microcavity sample~\cite{Cilibrizzi2014}, to verify the method using real experimental results as input to train the FNO model. Thus, the architecture of the FNO model coupled with the extra reservoir-scattering process is studied with numerical datasets and we demonstrate the first direct application of Neural Operators to a coupled exciton-polariton condensate system using an experimental dataset to train the algorithm. This work not only addresses the computational challenges of large-scale polariton simulations but also lays the groundwork for scalable workflows in the design of reconfigurable, all-optical devices.

\section*{Results}
\label{Section:Results}
\bmhead{Steady-state condensates}
\begin{figure*}[htbp]
\begin{center}
\includegraphics[scale=0.85]{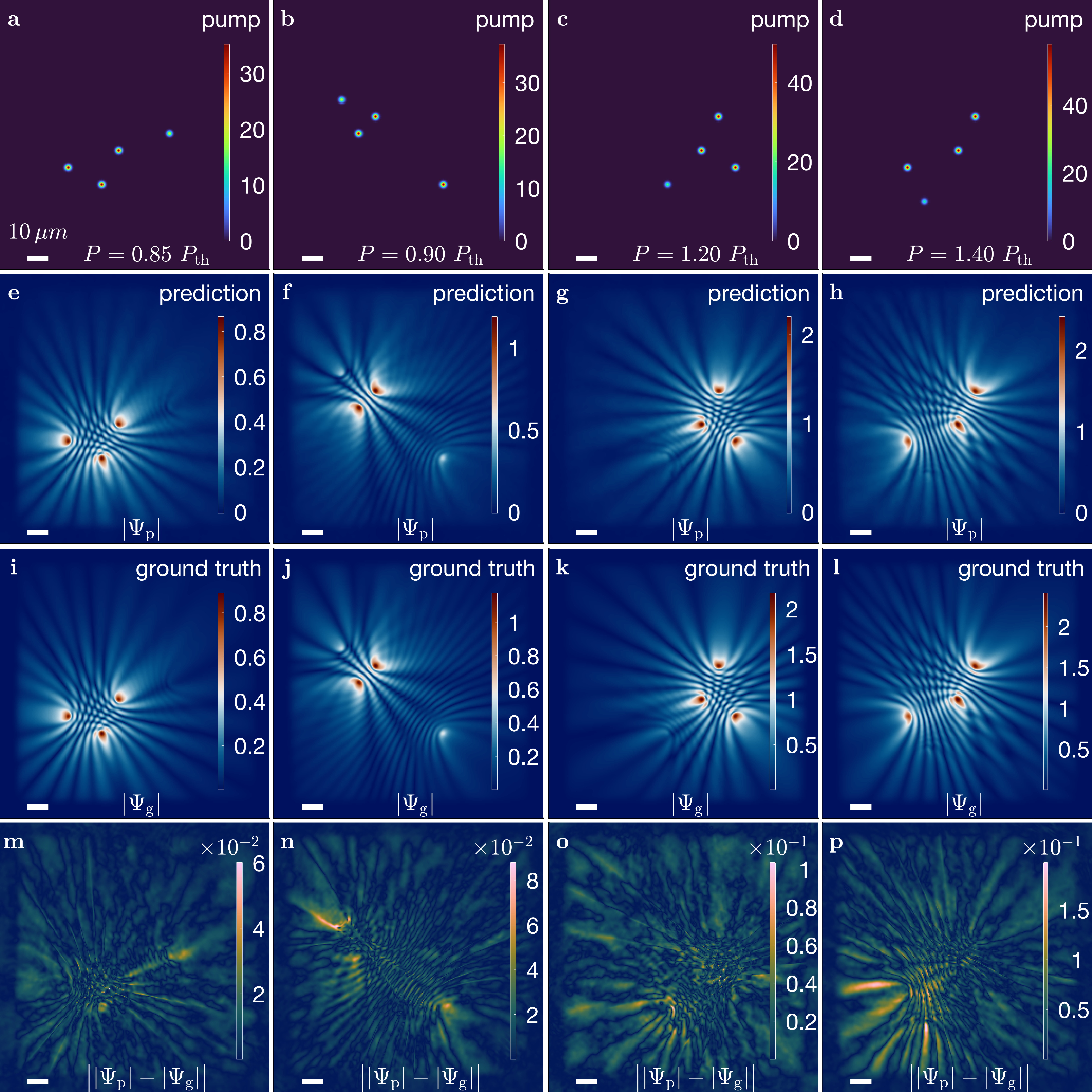}
\caption{\label{prediction_fno} \textbf{Comparison of the prediction with different short-distanced-spots pump configurations using theoretical datasets and the Fourier Neural Operator approach.} \textbf{a}-\textbf{d} From left to right, the different pump configurations are $P=0.85, 0.9, 1.2, 1.4\,P_{\mathrm{th}}$. \textbf{e}-\textbf{h} Corresponding condensate solutions $|\Psi_p|$ with pump profiles, each featuring a distinct spatial profile and intensity from the prediction datasets. \textbf{i}-\textbf{l} Corresponding numerical steady-state solutions $|\Psi_g|$ from the ground truth. \textbf{m}-\textbf{p} Corresponding absolute errors between prediction and ground truth $\big||\Psi_p|-|\Psi_g|\big|$. The white bar on all panels is $10\,\mathrm{\mu m}$. The corresponding percentage errors of the number of condensate particles, taken from Fig.~\ref{S_curve}, are $1.12\%$, $4.07\%$, $0.04\%$, $0.24\%$. $P_{\mathrm{th}}$ is the threshold of the power density per single Gaussian spot.}
\end{center}
\end{figure*}

In Fig.~\ref{prediction_fno}, we present the predictions of the FNO model for $4$ representative test cases and their corresponding ground truths. Note that we have carefully chosen four distinct pump profiles (see Figs.~\ref{prediction_fno}\textbf{a}-\textbf{d}) to visualize the performance of the model with varying inputs. As we see in Figs.~\ref{prediction_fno}\textbf{e}-\textbf{h} for predictions and in Figs.~\ref{prediction_fno}\textbf{i}-\textbf{l} for ground truth, the model is highly accurate (see Figs.~\ref{prediction_fno}\textbf{m}-\textbf{p}) in predicting the steady-state solution to the GPE coupled with rate equation (see Methods). The FNO predictions demonstrate excellent agreement with numerical ground truth, capturing key features of the condensate density, including interference patterns and fringe parity. Notably, the model accurately predicts the direction of ballistic flow and scattering on below-threshold barriers, aligning with similar experimental results observed in inorganic semiconductor materials, where clear interference patterns have been reported~\cite{Toepfer2020, Toepfer2021}. We see that the predictions and the simulation ground truths are almost the same for different pump configurations, including the parity of fringes among spots. The parity of these fringes is responsive to the distance between spots~\cite{Toepfer2020}, which also indicates that our model is capable of capturing these details, such as the type of interaction between condensates. 

The error panels in Figs.~\ref{prediction_fno}\textbf{m}-\textbf{p} reveal that the highest discrepancies occur near pump locations. These deviations arise from the inherent nonlinearity of the system in these regions and the information loss caused by Fast Fourier Transform cut-off modes in the FNO architecture. However, errors outside the pump regions are minor, primarily attributable to nonlinear interactions between condensates. Empirically, the lower errors correspond to pump configurations where the distance between the pumps is smaller, leading to better interference predictions. Higher errors correspond to pump configurations where at least one pump is far from the other two, leading to worse interference pattern predictions. This is evident in the results where the pumps are very far apart (see Fig.~\ref{S_prediction_fno} in SI), compared to Fig.~\ref{prediction_fno} where pumps are closer together. It is worth mentioning that due to extra interaction, despite for pump configuration with power density being below the threshold for each Gaussion spot denoted as $P_{\mathrm{th}}$ (see Methods for details), as shown in Figs.~\ref{prediction_fno}\textbf{a} and~\ref{prediction_fno}\textbf{b}, the whole system is still above the threshold. Also, it is worthy noting that the time to predict numerical solutions for $1122$ cases (see Methods) using the CUDA-based numerical method took $3.35\times10^{4}\,\mathrm{s}$, while the FNO model took $\SI{8.78}{s}$.
\bmhead{S-curve of condensate particles}
\begin{figure*}[htbp]
\begin{center}
\includegraphics[scale=0.60]{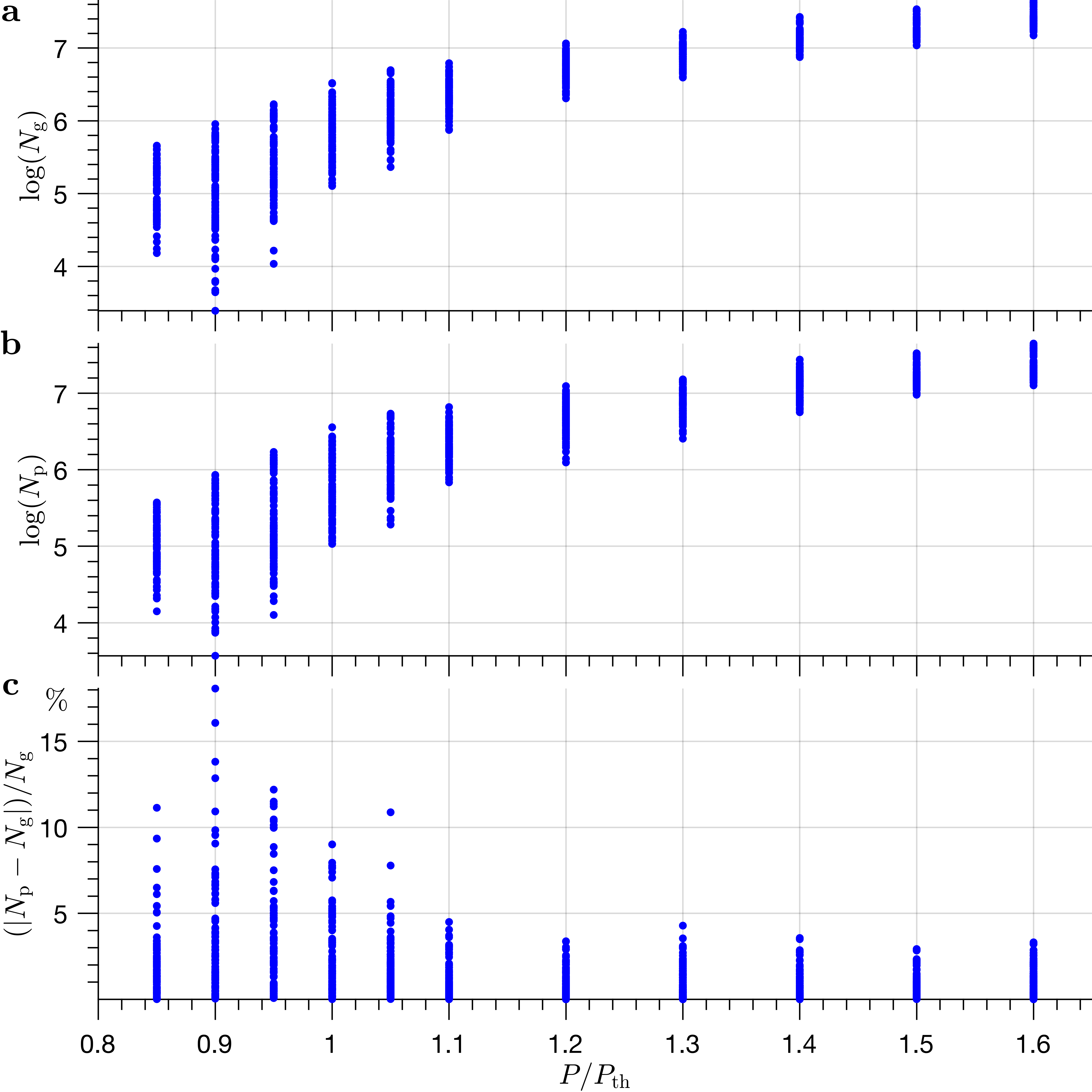}
\caption{\label{S_curve} \textbf{S-curve of the condensate particles as a function of the pumping density.} The logarithmic scale of number of particles for \textbf{a} ground truths denoted as $\mathrm{log}(N_{g})$,\textbf{b} predictions denoted as $\mathrm{log}(N_{p})$, and \textbf{c} the relative error of the condensate particles in the prediction $N_{p}$ with respect to the ground truth $N_{g}$ as a function of pumping density in the unit of $P_{\mathrm{th}}$, where $P_{\mathrm{th}}$ is the threshold power of a single Gaussian spot.}
\end{center}
\end{figure*}
As the system reaches the condensate threshold, the occupation of state will increase non-linearly then, followed by a linear increase as the excitation densities keep increasing~\cite{Kasprzak2006}, which is known as the S-curve of the condensate system. Here, we demonstrate that the robustness of the FNO model is that it works not only for the linear region when the pumping density is high but also for the weakly pumped region. In Fig.~\ref{S_curve}\textbf{a}, we plot the logarithmic scale of condensate particle numbers as a function of pump density for the ground truth, while Fig.~\ref{S_curve}\textbf{b} presents the FNO predictions. Both curves align closely, capturing the transition from weak to strong excitation regimes. The absolute relative errors of number of condensate particles from prediction and numerical solutions with respect to the numerical ones are shown in Fig.~\ref{S_curve}\textbf{c}. Most errors are less than $10\%$, which shows great robustness and consistency for the FNO model. Large errors below condensation threshold are expected because there are fewer training datasets with small numbers of particles and therefore show more errors in predictions; therefore, it shows more errors for the predictions. We can see that in Fig.~\ref{S_curve}\textbf{a} at $P=0.9\,P_{\mathrm{th}}$ there are more test cases with much lower particle numbers.
\bmhead{Experimental realization}
\begin{figure*}[htbp]
\begin{center}
\includegraphics[scale=0.85]{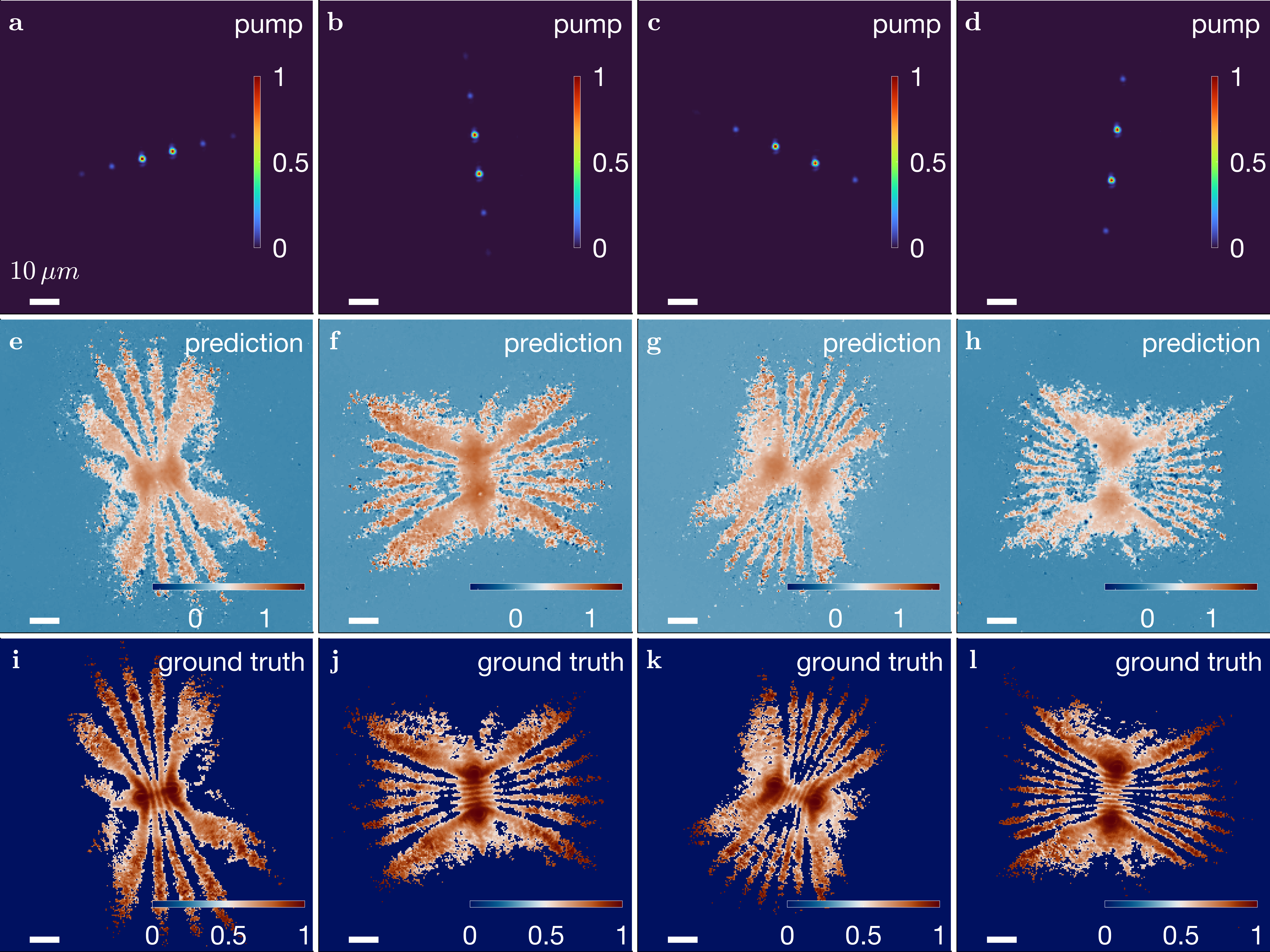}
\caption{\label{prediction_fno_experiment} \textbf{Comparison of the prediction with different normalized pump configurations using the preprocessed experimental datasets and Fourier Neural Operator approach.} \textbf{a}-\textbf{d} From left to right, the different pump configurations. \textbf{e}-\textbf{h} Corresponding predictions from the pump profiles.\textbf{i}-\textbf{l} Corresponding post-processed photoluminescence from the experiment. The number of fringes on \textbf{e}-\textbf{h} is $3$, $5$, $6$, $8$, respectively, which is the same as those on \textbf{i}-\textbf{l}. The white bar on all panels is $10\,\mathrm{\mu m}$. The pump density for the whole experiment is $3.6$ times the threshold value.}
\end{center}
\end{figure*}
 
Figure~\ref{prediction_fno_experiment} demonstrates the emission profile predictions obtained using experimental data as a training data set compared to the emission pattern obtained directly from the experiment. To obtain the desired spatial geometry of the pump profile, we calculate a series of spatial phase maps (kinoforms)~\cite{Lesem:IBM1969}. A feature of this method is the creation of additional weak spots aligned with the interaction axis of the main spots. They have a much lower intensity than the desired pump spots and thus do not cause the formation of unwanted condensates, and their interactions with the investigated condensates are negligible. All pump spots of the inputs shown in Figs.~\ref{prediction_fno_experiment}\textbf{a}-\textbf{d} are set to be equal at P = $3.6P_{\mathrm{th}}$. The number of fringes predicted from the FNO model, as shown in Figs.~\ref{prediction_fno_experiment}\textbf{e}-\textbf{h}, is $3$, $5$, $6$ and $8$. Even and odd parity indicates the antiferromagnetic and ferromagnetic order in the polariton dyad, respectively. The FNO model reproduces the spatial profile of emissions with high accuracy regardless of the type of interaction between condensates, which is confirmed by Figs.~\ref{prediction_fno_experiment}\textbf{i}-\textbf{l} presenting experimental emission profiles. The effectiveness of the method was confirmed by an accurate reconstruction of the emission pattern with the correct number of interference fringes and propagation trajectories of polaritons compared to the ground truth obtained from the experiment. Moreover, the comparison of Figs.~\ref{prediction_fno_experiment}\textbf{e}-\textbf{h} and Figs.~\ref{prediction_fno_experiment}\textbf{i}-\textbf{l} shows that even subtle details of the patterns such as local intensity minima and maxima of the intensity profiles have been reproduced correctly. The method of post-processing of the experimental data is detailed in SI. The details of the experimental datasets and hyperparameters can be found in Methods and SI, respectively. More prediction results with different pump profiles can be found in SI.
\section*{Discussion}
Various general ML methods have been proposed to incorporate the underlying physics-based losses and information into the model to aid the learning task, such as in \cite{Raissi2019,Karniadakis2021,Lu2021,Cuomo2022}. In this work, we have taken a purely data-driven approach to training; however, we believe that incorporating additional physics-informed loss terms will strictly increase the rate of convergence and accuracy. This is especially appealing, given that we have a strong theoretical understanding of the underlying system. In contrast to the theoretical steady-state datasets, the time-integrated PL data can also achieve good agreement with experimental features. A similar FNO-based real-world data-driven treatment has been adopted for weather forecasts~\cite{pathak2022fourcastnet}. In contrast to theoretical datasets, preprocessing of the experimental datasets is critical, as the input parameters from the experimental devices usually come with different orders of magnitude of values. It is important to note that the prediction from the experimental pump profile deviates slightly from the uniform values of the ground truth. Since only the relative intensity of the PL matters, it is not an issue from a physics perspective. Moreover, with the help of a streak camera, PL can be captured at the picosecond level, making it possible to make predictions of a time-resolved condensate formation on the basis of purely experimental data.

In summary, we explored the potential of the FNO in the context of polariton condensates. Our findings demonstrate a notable alignment with the simulation data, with an approximate three orders of magnitude speed up in solution generation compared to CUDA-based GPU solvers. This research not only paves the way for the conceptualization and development of advanced large-scale all-optical devices from both theoretical and experimental perspectives but also draws parallels with the principles of EDA traditionally used in chip design. This approach represents a significant step toward the development of scalable workflows for designing reconfigurable optical devices.

\section*{Methods}

\bmhead{Gross-Pitaevskii equation}
%Polaritons, often described as quasiparticles with half-light and half-matter characteristics, have an impressively low effective mass due to their photonic components. This mass is approximately five orders of magnitude less than that of a bare electron. Thus, the necessary temperature for condensation is approximately $\SI{10}{K}$ for inorganic semiconductor materials~\cite{Kasprzak2006}, which contrasts significantly with atomic condensates such as Rubidium-87, which require temperatures around \SI{170}{nK}~\cite{Anderson1995}. Notably, polariton condensation can be realized using organic materials even at room temperature~\cite{Christopoulos_PRL2007, sanvitto2016road, su2018room}. In polariton condensates, due to their excitonic component, the predominant repulsive interaction between polaritons results in a blue-shifting effect on the potential and gives rise to rich nonlinear effects~\cite{Wertz2010, Schmutzler2014}. In essence, it is the photonic (light-like) component of the polaritons that confers their notably light effective mass suitable for condensates, while the excitonic (matter-like) component is responsible for the observed nonlinear effects. 

The dynamics of polariton condensates are governed by the Gross-Pitaevskii equation (GPE) coupled with the rate equation of the exciton reservoir $\mathcal{N}$~\cite{Wouters2007}:
\begin{gather}
i\hbar\frac{\partial }{\partial t}\Psi=\bigg[-\frac{\hbar^{2}}{2m}\nabla^{2}
 +\alpha |\Psi|^2 +G\Big(\mathcal{N} + \frac{\eta }{\Gamma}P(\bm{r})\Big)
+i\frac{ \hbar}{2}\big[R \mathcal{N}-\gamma\big]\bigg]\Psi,
\label{Gross-Pitaevskii-equation}
\\
\frac{\partial }{\partial t}\mathcal{N}=-\Big[\Gamma+R |\Psi|^{2}\Big]\mathcal{N} + P(\bm{r}),
\label{reservoir-rate-equation}
\end{gather}
where $m$ is the polariton effective mass, $\alpha$ and $G$ stand for, respectively, polariton-polariton and polariton-reservoir interaction, $R$ denotes the scattering rate from the reservoir to the condensates, $\eta$ refers to the ratio of the dark excitons, and $\gamma$ ($\Gamma$) is the decay rate of the polariton (reservoir). The detuning between the exciton and the photon mode can greatly alter the interaction terms with the relationship $\alpha=g|\chi|^4$ and $G=2g|\chi|^2$, and $g=g_{0}/N$ where $g_{0}$ is the exciton-exciton interaction, $N_{\text{QW}}$ is the number of QWs, and $|\chi|^2$, representing the percentage of exciton of which the polariton consists, is the Hopfield coefficient~\cite{Hopfield1958} of the excitonic branch. In this work, the continuous-wave (CW) pump, denoted by $P(\bm{r})$, is used to replenish the reservoir, which is depleting due to the dissipative character of the polaritonic system. The nonlinear term $|\psi|^{2}$ appearing in both the pump-to-reservoir transition (see~\eqref{reservoir-rate-equation}) and the superfluid in the condensates (see~\eqref{Gross-Pitaevskii-equation}), produce the rich nonlinear characteristic induced from the pump to the condensate. 

In the case of CW excitation under a weak pumping regime, the approximate value of $|\psi|^2$ tends towards zero. In this situation, the rate of reservoir with respect to time maintains a stationary state, or in mathematical terms, $\partial \mathcal{N}/\partial t=0$. The determination of threshold power, denoted at $P_{\mathrm{th}}$, is possible through an analysis of the right-hand side (r.h.s.) of \eqref{Gross-Pitaevskii-equation} where $R\mathcal{N}=\gamma$ serves as a representative of the equilibrium state between gain and loss. Therefore, the threshold power $P_{\mathrm{th}}=\gamma\Gamma/R$ is obtained. This suggests that when the population of polaritons exceeds the condensation threshold $P_{\mathrm{th}}$, a detectable density value manifests itself. The real potential of~\eqref{Gross-Pitaevskii-equation} denoted $V$ in the stationary state of the system, therefore, is
\begin{gather}
V(\bm{r}) = \alpha|\psi|^2 + G\Big(\frac{1}{\Gamma + R|\psi|^2} + \frac{\eta }{\Gamma}\Big)P(\bm{r}).
\label{real-potential}
\end{gather}
The real potential is composed of two main components: one originating from the pumping region (first term on the r.h.s of~\eqref{real-potential}) and the other stemming from the interactions among the polaritons outside this region (second term on the r.h.s of~\eqref{real-potential}). When the pump power is below the threshold, the direct contribution of the potential goes directly into the pumping profile. This relationship is represented as $V(\bm{r}) = (1+\eta)(G/\Gamma)P(\bm{r})$. The spatial profile is chosen for the demonstration of $N_{G}$ Gaussian spots. That is 
\begin{gather}
P(\bm{r}) = \sum_{i}^{N_{G}}P_{i}G_{i}(\bm{r}),
\label{pump-profile}
\end{gather}
where $P_{i}$ stands for strength of each spot and the normalized Gaussian function $G_{i}(\bm{r})$, with full width at half maximum (FWHM) denoted $\sigma$, is defined as
\begin{gather}
G_{i}(\bm{r}) = \frac{1}{2\pi \sigma^2}\exp{\bigg(\frac{-|\bm{r}-\bm{r}_{i}|}{2\sigma^2}\bigg)}.
\label{Gaussian-function}
\end{gather}
Note that $\bm{r}_{i}$ represents a different location of spots. 

The FNO prediction shown in Section~\ref{Section:Results} takes $P(\bm{r})$ and $|\Psi(t=0)|$ as two inputs. Here, $|\Psi(t=0)|$ is fixed as a zero matrix with the same dimension as the matrix of $P(\bm{r})$. The number of condensate particles in Fig.~\ref{S_curve} are calculated from
\begin{gather}
N = \int |\Psi(\bm{r})|^2 d^{2}\bm{r}.
\label{number_condensate_particles}
\end{gather}

%%%%%%%%%%%%%%%%%%%%%%%%%%%%%%%%%
\bmhead{Fourier Neural Operators}
\label{Sec:Fourier-Neural-Operator}
\begin{figure*}[htbp]
\centering
\begin{tikzpicture}[node distance=0.7cm and 0.5cm,
                    >={Stealth[round, scale=1.2]}] % Adjust the arrow tip style here
    % Define styles
    \tikzstyle{circleNode} = [circle, minimum size=1cm, draw=black, thick, fill=blue!40, text=white]
    \tikzstyle{rectNode} = [rectangle, minimum width=1.5cm, minimum height=0.7cm, draw=black, thick, fill=red!30, text=black]
    \tikzstyle{largeBox} = [rectangle, minimum width=6.5cm, minimum height=5.0cm, draw=black, thick, fill=orange!20, opacity=0.7]
    \tikzstyle{innerBox} = [rectangle, minimum width=6cm, minimum height=3.0cm, draw=black, thick, fill=green!30, opacity=0.8]

    % Draw nodes
    \node[circleNode] (c1) {a(x)};
    \node[circleNode, right=of c1] (c2) {$\mathcal{P}$};
    \node[rectNode, right=of c2] (r1) {Fourier layer};
    \node[right=of r1, font=\Large] (dots) {$\cdots$};
    \node[rectNode, right=of dots] (r4) {Fourier layer};
    \node[circleNode, right=of r4] (c3) {$\mathcal{Q}$};
    \node[circleNode, right=of c3] (c4) {u(x)};
    \node[largeBox, below=of r1] (large) {};
    \node[circleNode, at=(large.center), shift={(-2.3cm,-1.80cm)}] (boxC1) {$\nu(x)$};
    \node[circleNode, right=of boxC1] (boxC2) {$W$};
    \node[circleNode, right=of boxC2] (boxC3) {$+$};
    \node[circleNode, right=of boxC3] (boxC4) {$\sigma$};
    \node[innerBox, below=of large.north, xshift = 0.0 cm, yshift=0.5cm] (inner) {};

    % Add three circles inside the smaller box
    \node[circleNode, at=(inner.center), shift={(-2.25cm,0cm)}] (innerC1) {$\mathcal{F}$};
    \node[circleNode, at=(inner.center)] (innerC2) {$R$};
    \node[circleNode, at=(inner.center), shift={(2.25cm,0cm)}] (innerC3) {$\mathcal{F}^{-1}$};

    % Draw multiple frequency bands between \mathcal{F} and R
    \foreach \i in {0.7,0.5,0.3, -0.5,-0.7,-0.9, -1.1, -1.3} {
        \draw[thick] ($(innerC1.east) + (0,\i)$) -- ($(innerC2.west) + (0,\i)$);
    }
    \node[font=\LARGE] at ($(innerC1.east)!0.5!(innerC2.west)$) {$\vdots$};
    \node[anchor=south,align=center] at ($(innerC1.east) + (0.6,0.65)$) {Multiple \\ modes};

    % Draw multiple frequency bands between R and \mathcal{F}^{-1}
    \foreach \i in {0.5,0.3, -0.5,-0.7,-0.9} {
        \draw[thick] ($(innerC2.east) + (0,\i)$) -- ($(innerC3.west) + (0,\i)$);
    }
    \node[font=\LARGE] at ($(innerC2.east)!0.5!(innerC3.west)$) {$\vdots$};
    \node[anchor=south,align=center] at ($(innerC2.east) + (0.6,0.8)$) {128 modes};

    % Connect nodes
    \draw[->, thick] (c1) -- (c2);
    \draw[->, thick] (c2) -- (r1);
    \draw[->, thick] (r1) -- (dots);
    \draw[->, thick] (dots) -- (r4);
    \draw[->, thick] (r4) -- (c3);
    \draw[->, thick] (c3) -- (c4);
    \draw[dashed, thick] (r1.south west) -- (large.north west);
    \draw[dashed, thick] (r1.south east) -- (large.north east);
    \draw[->, thick] (boxC1) -- (inner.south west);
    \draw[->, thick] (inner.south east) -- (boxC3);
    \draw[->, thick] (boxC1) -- (boxC2);
    \draw[->, thick] (boxC2) -- (boxC3);
    \draw[->, thick] (boxC3) -- (boxC4);

    % Flourish/brace directly connecting to the upper corners
    \draw [decoration={brace, amplitude=10pt, raise=4pt}, decorate, thick]
        (r1.north west) -- node[above=14pt] {4} (r4.north east);
\end{tikzpicture}
\caption{\label{Architecture_FNO} \textbf{Architecture of the Fourier Neural Operator.} The process begins with the input $a(x)$ which undergoes a lifting operation, denoted as $\mathcal{P}$. This is followed by $4$ consecutive Fourier layers. Subsequently, a projector $\mathcal{Q}$ transforms the data to the desired target dimension, resulting in the output $u(x)$. The inset provides a detailed view of the structure of a Fourier layer. Data initially flow to the layer as $\nu(x)$ and are bifurcated into two branches: one undergoes a linear transformation $W$, and the other first experiences a Fourier transformation, from which the 128 lowest Fourier modes are kept, and the other higher modes are filtered out by undergoing a transformation $R$, and ends with an inverse Fourier transformation with these left modes. The two data streams then converge, followed by the application of an activation function $\sigma$.}
\end{figure*}
The numerical solution to \eqref{Gross-Pitaevskii-equation} and~\eqref{reservoir-rate-equation} is derived using the SSFM, detailed in Supplementary Information (SI). A natural ML analog to this classical method is the FNO architecture~\cite{li2021fourier}. More generally, Neural Operators~\cite{kovachki2023neural} are a class of models which learn mappings between two infinite-dimensional spaces from a finite set of input-output pairs. Many variants of the Neural Operator architecture have been applied to approximate solutions to Partial Differential Equations, such as in~\cite{gopakumar2023fourier, WEN2022104180, 10.1190/geo2022-0268.1, LI2022100389}. The Neural Operator architecture consists of a lifting operation $\mathcal{P}$, followed by iterative updates using a Kernel Integral Operator $\mathcal{K}$, and a final projection operator $\mathcal{Q}$, as defined in~\eqref{eq:Neural-Operator}.
\begin{gather}
\label{eq:Neural-Operator}
        \mathcal{G}_\theta := \mathcal{Q} \circ \sigma_T(W_{T-1} + \mathcal{K}_{T-1} + b_{T-1}) \circ ... \circ \sigma_1(W_0 + \mathcal{K}_0 + b_0) \circ \mathcal{P}
\end{gather}
Here, $\sigma$ corresponds to a non-linearity and $W$ and $b$ correspond to the weights and biases of the Kernel Integral Layer, respectively. $\mathcal{P}$ and $\mathcal{Q}$ are point-wise fully local projection and lifting operators. The choice of the Kernel Integral Operator $\mathcal{K}$ delineates the class of the Neural Operator. Specifically, the FNO (see Fig.~\ref{Architecture_FNO}) uses the Kernel Integral Operator defined by:
\begin{gather}
\label{eq:Fourier-Convolution-Operator}
        (\mathcal{K}_t(v_t))(x) = \mathcal{F}^{-1}(\mathcal{R}_\phi \cdot \mathcal{F}(v_{t-1}))(x) \quad\quad \forall x \in \mathbb{R}^n
\end{gather}
Here $\mathcal{F}$ and $\mathcal{F}^{-1}$ correspond to the Fourier and Inverse Fourier Transforms and $\mathcal{R}_\phi$ corresponds to the Fourier Transform of a periodic function arising from the definition of a Kernel Integral Operator given in \cite{li2021fourier}. This object is parameterised by a linear transformation of the top $k$ modes pertaining to the given layer, which acts as a hyperparameter in the model.

The natural choice of the FNO architecture for approximating the solution to \eqref{Gross-Pitaevskii-equation} and~\eqref{reservoir-rate-equation} is due to the inductive bias that arises from the SSFM - FNO correspondence stated below.

\begin{theorem} \textbf{(SSFM-FNO Correspondence)}
\label{Thm:SSFM-FNO-Correspondence}
\textit{Suppose that $\sigma \in (TW)$ is a Tauber-Wiener function, $X$ is a Banach Space, $K \subset X$ is a compact set, $V$ is a compact set in $C(K)$, $\Psi_t$ is a nonlinear continuous operator representing the solution of the first-order Split-step Fourier Method at time $t$. Then for any any $\epsilon > 0$, there are a positive integer $n$, $m$ points $x_1,..., x_m \in K$, and real constants $c_i$, $\theta_i$, $\xi_{ij}$ (for $i=1,...,n$ and $j=1,...,m$) such that:}
\begin{gather}
\label{FNO-R}
\mathcal{R}_\phi := \sum_{i=1}^n c_i \sigma \left( \sum_{j=1}^m \xi_{ij} u(x_j) + \theta_i \right),
\end{gather}
\begin{gather}
\label{FNO-Theorem}
\Biggl| \Psi_{t+\Delta t}(\Psi_t)(x) - \mathcal{F}^{-1}(\mathcal{R}_\phi \cdot \mathcal{F}(v_t))(x) ) \Biggr| < \epsilon
\end{gather}
\textit{holds for all $u \in V$.}
% \end{theorem}
\end{theorem}
\textit{Proof.} See SI.
\bmhead{Sample and experimental techniques}
The sample used in the experiment is the 2$\lambda$ high-quality semiconductor optical microcavity with quantum wells~\cite{Cilibrizzi2014}. The structure consists of a GaAs-based microcavity placed between two DBRs made of pairs of GaAs and AlAs$_{0.98}$P$_{0.02}$ layers. In the microcavity region, the three pairs of $\SI{6}{nm}$ In$_{0.08}$Ga$_{0.92}$As QWs placed in anti-nodes of the electric field. Two additional QWs positioned at the extreme nodes of the cavity wells serve for carrier collection. The sample was held in a cold finger, closed-cycle cryostat operating at a temperature of $T\approx\SI{7}{K}$. 

The optical nonresonant excitation is provided by a continuous-wave
Ti:Sapphire laser modulated by an acousto-optic modulator to prevent heating effects. In order to obtain the pump profile with multiple-spot excitation, a reflecting liquid-crystal spatial light modulator (SLM) is used. The screen of the SLM displays calculated phase holograms in the Fourier plane modulating the Gaussian beam of the excitation laser beam. The phase holograms are accomplished by imprinting an analytically generated phase pattern on the SLM screen. The procedure results in generating the intended configuration of the laser spots at the focal plane of the microscope objective lens.
%%%%%%%%%%%%%%%%%%%%%%%%%%%%%%%%%%%%
%%%%%%%%%%%%%%%%%%%%%%%%%%%%%%%%%%%%
%%%%%%%%%%%%%%%%%%%%%%%%%%%%%%%%%%%%
\bmhead{Numerical simulation}
To better emulate the experiment, $\sigma\approx 0.85\,\mathrm{\mu m}$, the FWHM of each Gaussian spot equalling to $2\,\mathrm{\mu m}$, is chosen. The simulation is based on InGaAs QWs~\cite{Cilibrizzi2014} with slightly negatively detuned cavities. The parameters are the following: $m=0.28\mathrm{\,meV\,ps^{2}\,\mu m^{-2}}$, $|\chi|^{2}=0.4$, $N_{\text{QW}}=6$, $g_{0}= 0.01 \,\mathrm{meV\,\mu m^{2}}$, $\hbar R=10g$, $\eta=2$, and $\gamma^{-1} = \Gamma^{-1} = \SI{5.5}{ps}$.
%%%%%%%%%%%%%%%%%%%%%%%%%%%%%%%%%%%%
%%%%%%%%%%%%%%%%%%%%%%%%%%%%%%%%%%%%
%%%%%%%%%%%%%%%%%%%%%%%%%%%%%%%%%%%%
\bmhead{Numerical dataset}
The datasets are constructed based on varying pump profiles $P(\bm{r})$, as described in~\eqref{pump-profile}. This profile is characterized by four Gaussian spots taking $N_{G}=4$ with the spatial profile of each spot obtained by $G_{i}(\bm{r})$. Among these four spots, three of them are equally powered and have their power set at $P_{i}=0.85, 0.90, 0.95, 1.0, 1.05, 1.1, 1.2, 1.3, 1.4, 1.5, 1.6\,P_{\mathrm{th}}$, while the fourth spot is powered far below the threshold at $P_{i}=0.5\,P_{\mathrm{th}}$. Thus, in terms of the power value, there are $11$ different configurations. Note that $P_{\mathrm{th}}$ refers to the threshold power for a single spot, which means that the entire system can still trigger the condensate with three spots below the threshold with an additional contribution from the interaction among them~\cite{Tosi2012}. The reason why the power varies is that we want the datasets to also cover the S-curve (see Fig.~\ref{S_curve}\textbf{a}), the region where the power-intensity relationship~\cite{Kasprzak2006} is taken into account. These profiles are stochastically determined within a square region that measures $64\,\mathrm{\mu m} \times 64\,\mathrm{\mu m}$ out of the entire configuration with $128\,\mathrm{\mu m} \times 128\,\mathrm{\mu m}$. The region where Gaussian spots stay is smaller than the full grid, to make sure that they are still far from the region where the boundary condition is applied. Care has also been taken to ensure that the Gaussian spots do not overlap under the same power density, so that without losing generality the minimal distance between two spots is set at $4\times\mathrm{FWHM}$ of the Gaussian spot. Additionally, every pump profile is unique, and among the spots with power exceeding the threshold, each one is distinct from the others, thereby eliminating any potential redundancy. Given $0.5\,\mathrm{\mu m}$ resolution per pixel per dimension, the total datasets for the pump configuration is with size $256\times256\times11220$ where $256$ represents each square map size per dimension and $11220$ is the number of different pump configurations (of which $1122$ configurations are used for testing, $1122$ configurations are used for validation, and $8976$ for training respectively). The datasets for the density map are of size $256\times256\times2\times11220$ where $2$ refers to the density at the initial and final time. It is worth mentioning that the systems of all the datasets are chosen with a system only at stationary state with single energy mode, which means that the results with multiple energy modes are excluded. In multimode cases, the wavefunction density changes at different times, which can be found in experiments~\cite{Krizhanovskii2009, Toepfer2020}.
%%%%%%%%%%%%%%%%%%%%%%%%%%%%%%%%%%%%
%%%%%%%%%%%%%%%%%%%%%%%%%%%%%%%%%%%%
%%%%%%%%%%%%%%%%%%%%%%%%%%%%%%%%%%%%
\bmhead{Experimental dataset}
The experimental pump profiles are normalized to unity. PL data are enhanced using logarithmic function and contrast-limited adaptive histogram equalization~\cite{zuiderveld1994contrast}, which is detailed in SI. The datasets are of size $256\times256\times1120$ of which $1104$ cases are used for training and $16$ cases for testing. The initial state is a zero-valued array of size $256\times256\times1120$. Data argumentation is applied for the training datasets by rotating the original $138$ training datasets at $45^{\circ}$ step around the image center, namely, $0^{\circ}$, $45^{\circ}$, $90^{\circ}$, $\dots$, $315^{\circ}$, at the center of the image, resulting in training datasets of size $1104$.
\section*{Acknowledgments}
This work was supported by the European Union Horizon 2020 program, through a Future and Emerging Technologies (FET) Open research and innovation action under Grant Agreement No. 964770 (TopoLight).

\section*{Contributions}

 \section*{Competing interests.} 

Authors declare that they have no competing interests.    

 \section*{Data and Materials Availability.} 
 
All data supporting this study are openly available from the University of Southampton repository [link is to be provided].

%% BioMed_Central_Bib_Style_v1.01

\bibliography{sn-bibliography}

\setcounter{equation}{0}
\setcounter{figure}{0}
\setcounter{section}{0}
\renewcommand{\theequation}{S\arabic{equation}}
\renewcommand{\thefigure}{S\arabic{figure}}
\renewcommand{\thesection}{S\arabic{section}}

\newpage
\vspace{1cm}
\begin{center}
\Large \textbf{Supplementary Information}
\end{center}

\section{Split-step Fourier method}
\label{App:Split-step-method}
In this Appendix, the numerical solution to~\eqref{Gross-Pitaevskii-equation} and~\eqref{reservoir-rate-equation} from the main content is derived using split-step Fourier method (SSFM). \eqref{Gross-Pitaevskii-equation} from the main content can rearranged as
\begin{gather}
\frac{\partial }{\partial t}\Psi=\bigg\{i\frac{\hbar}{2m}\nabla^{2}
 -i\frac{\alpha}{\hbar} |\Psi|^2 -i\frac{G}{\hbar}\Big[\mathcal{N} + \frac{\eta }{\Gamma}P\Big]
+\frac{ 1}{2}\big[R \mathcal{N}-\gamma\big]\bigg\}\Psi.
\label{rearranged-GPE}
\end{gather}
The direct solution from~\eqref{rearranged-GPE} at time interval $[t, t+\Delta t]$ is
\begin{gather}
\Psi_{t+\Delta t}=\exp{[(\widehat{f}_{L}+\widehat{f}_{N})\Delta t]}\Psi_{t},
\end{gather}
where 
\begin{gather}
\widehat{f}_{L} = i\frac{\hbar}{2m}\nabla^{2},
\\
\widehat{f}_{N} =  -i\frac{\alpha}{\hbar} |\Psi|^2 -i\frac{G}{\hbar}\Big[\mathcal{N} + \frac{\eta }{\Gamma}P\Big]
+\frac{ 1}{2}\big[R \mathcal{N}-\gamma\big].
\end{gather}
Note that since $[\widehat{f}_{L},\widehat{f}_{N}]\neq 0$, the exponentiation identity cannot be applied directly, namely, $\exp{[\widehat{f}_{L}+\widehat{f}_{N}]}\neq\exp{[\widehat{f}_{L}]}\exp{[\widehat{f}_{N}]}$. By applying the Baker–Campbell–Hausdorff (BCH) formula at second order and the strang splitting, we have
\begin{gather}
\Psi_{t+\Delta t} = \exp{[\frac{1}{2}\widehat{f}_{N}\Delta t]}\exp{[\widehat{f}_{L}\Delta t]}\exp{[\frac{1}{2}\widehat{f}_{N}\Delta t]}\Psi_{t},
\label{psi-operator-expression}
\end{gather}
which give the accuracy of $\Delta t^{3}$. The essence of SSFM is that it can convert the nonlinear operator $\widehat{f}_{N}$ into the linear one through the Fourier transform. We can always construct the relation that
\begin{gather}
\frac{\partial}{\partial t}\Psi=\widehat{f}_{L}\Psi.
\label{construct-linear-operator}
\end{gather}
Then, applying the Fourier transform for above equation, we have
\begin{gather}
\frac{\partial}{\partial t}\mathcal{F}[\Psi]=\widehat{f}_{P}\mathcal{F}[\Psi],
\end{gather}
where
\begin{gather}
\widehat{f}_{P} = -i\frac{\hbar}{2m}|\bm{k}|^2,
\end{gather}
is the momentum operator. This implies that~\eqref{construct-linear-operator} can be rewritten as
\begin{gather}
\Psi_{t+\Delta t} = \exp{[\frac{1}{2}\widehat{f}_{N}\Delta t]}\mathcal{F}^{-1}\bigg[\exp{[\widehat{f}_{P}\Delta t]}\mathcal{F}\Big[\exp{[\frac{1}{2}\widehat{f}_{N}\Delta t]}\Psi_{t}\Big]\bigg].
\label{full-psi-operator-expression}
\end{gather}
Ultimately, the final state wavefunction can be obtained by iteraction of~\eqref{full-psi-operator-expression} overall the infinitesimal time steps. During each iteration, $\mathcal{N}$ is updated within the operator $\widehat{f}_{N}$. That is,
\begin{gather}
\mathcal{N}_{t+\Delta t} = \exp{\big[(-\Gamma+R|\Psi|^2)\Delta t\big]}\mathcal{N}_{t}+P\Delta t.
\end{gather}
The kernel defined by~\eqref{psi-operator-expression} is used to calculate the ground truth for the simulation with BCH at second order. However, if we limit ourselves to BCH at first order, then the corresponding~\eqref{full-psi-operator-expression} can be simplified as
\begin{gather}
\Psi_{t+\Delta t} = \mathcal{F}^{-1}\bigg[\exp{[\widehat{f}_{P}\Delta t]}\mathcal{F}\Big[\exp{[\widehat{f}_{N}\Delta t]}\Psi_{t}\Big]\bigg].
\label{full-psi-operator-expression-first-order}
\end{gather}
From here, \eqref{full-psi-operator-expression-first-order} takes the functional form of the Kernel Integral Operator of the FNO introduced in Section~\ref{Sec:Fourier-Neural-Operator}. This is stated formally in the SSFM - FNO correspondence given in Theorem~\ref{Thm:SSFM-FNO-Correspondence}.

\setcounter{theorem}{0}
\setcounter{equation}{0}
\renewcommand{\theequation}{B\arabic{equation}}
\renewcommand{\thetheorem}{B\arabic{theorem}}
%%%%%%%%%%%%%%%%%%%%%%%%%%%%%%%%%
%%%%%%%%%%%%%%%%%%%%%%%%%%%%%%%%%
%%%%%%%%%%%%%%%%%%%%%%%%%%%%%%%%%
\section{Theorems}
\label{App:SSFM-FNO}
\begin{theorem}
\textbf{(Universal Approximation Theorem for Functionals.)}
\textit{Suppose that $\sigma \in (TW)$ is a Tauber-Wiener function, $X$ is a Banach Space, $K \subset X$ is a compact set, $V$ is a compact set in $C(K)$, $f$ is a continuous functional defined on $V$, then for any $\epsilon > 0$, there are a positive integer $n$, $m$ points $x_1,..., x_m \in K$, and real constants $c_i$, $\theta_i$, $\xi_{ij}$, $i = 1,..., n$, $j = 1,..., m$, such that}

\begin{gather}
\label{Universal-Functional-Theorem}
\left| f(u) - \sum_{i=1}^n c_i \sigma \left( \sum_{j=1}^m \xi_{ij} u(x_j) + \theta_i \right) \right| < \epsilon
\end{gather}

holds for all $u \in V$.

\end{theorem}
\textit{Proof.} See \cite{Lu_2021_DeepONet} and \cite{universal:3561150, Functionals92Chen, RBF95Chen}. 
\newline

\begin{theorem}
\textbf{(Universal Approximation Theorem for Operators.)}
\textit{Suppose that $\sigma \in (TW)$ is a Tauber-Wiener function, $X$ is a Banach Space, $K_1 \subset X$, $K_2 \subset \mathbb{R}^d$ are two compact sets in $X$ and $\mathbb{R}^d$, respectively, $V$ is a compact set in $C(K_1)$, $G$ is a nonlinear continuous operator, which maps $V$ into $C(K_2)$, then for any $\epsilon > 0$, there are positive integers $n$, $p$, $m$, constants $c^k_i$, $\xi^k_{ij}$, $\theta^k_i$, $\zeta_k \in \mathbb{R}$, $w_k \in \mathbb{R}^d$, $x_j \in K_1$, $i = 1,..., n$, $k = 1,..., p$, $j = 1,..., m$, such that}

\begin{gather}
\label{Universal-Operator-Theorem}
\left| G(u)(y) - \sum_{k=1}^p \sum_{i=1}^n c^k_i \sigma \left( \sum_{j=1}^m \xi^k_{ij} u(x_j) + \theta^k_i \right) \sigma(w_k \cdot y + \zeta_k ) \right| < \epsilon
\end{gather}

\textit{holds for all $u \in V$ and $y \in K_2$.}
\end{theorem}
\textit{Proof.} See \cite{Lu_2021_DeepONet} and \cite{universal:3561150, Universal95Chen, RBF95Chen}.  
\newline

\begin{theorem}
\textbf{SSFM-FNO Correspondence.}
\label{SSFM-FNO-Proof}
\textit{Suppose that $\sigma \in (TW)$ is a Tauber-Wiener function, $X$ is a Banach Space, $K \subset X$ is a compact set, $V$ is a compact set in $C(K)$, $\Psi_t$ is a nonlinear continuous operator representing the solution of the first-order Split-step Fourier Method at time $t$, then for any $\epsilon > 0$, there are a positive integer $n$, $m$ points $x_1,..., x_m \in K$, and real constants $c_i$, $\theta_i$, $\xi_{ij}$, $i = 1,..., n$, $j = 1,..., m$, such that for}
\begin{gather}
\mathcal{R}_\phi := \sum_{i=1}^n c_i \sigma \left( \sum_{j=1}^m \xi_{ij} u(x_j) + \theta_i \right),
\end{gather}
\begin{gather}
\Biggl| \Psi_{t+\Delta t}(\Psi_t)(x) - \mathcal{F}^{-1}(\mathcal{R}_\phi \cdot \mathcal{F}(v_t))(x) ) \Biggr| < \epsilon
\end{gather}
\textit{holds for all $u \in V$.}
\end{theorem}
\textit{Proof.} Limiting the solution of \eqref{Gross-Pitaevskii-equation} to first-order in the BCH formula, we get 
\begin{gather}
\Psi_{t+\Delta t}(\Psi_t)(x) = \mathcal{F}^{-1}\bigg[\exp{[\widehat{f}_{P}\Delta t]}\mathcal{F}\Big[\exp{[\widehat{f}_{N}\Delta t]}\Psi_{t}\Big]\bigg].
% \label{full-psi-operator-expression-first-order}
\end{gather}

Also considering the Kernel Integral Operator for the FNO as defined in~\eqref{eq:Fourier-Convolution-Operator}, we have
\begin{gather}
% \label{eq:Fourier-Convolution-Operator}
        (\mathcal{K}_{t+1}(v_t))(x) = \mathcal{F}^{-1}(\mathcal{R}_\phi \cdot \mathcal{F}(v_{t}))(x) \quad\quad \forall x \in \mathbb{R}^n
\end{gather}
It is clear that these two equations are equivalent for some $\mathcal{R}_\phi = \exp{ [\widehat{f}_P \Delta t]}$ and $v_t = \exp{[\widehat{f}_N \Delta t] \Psi_t}$.

Now, appealing to the Universal Approximation Theorems~\ref{Universal-Functional-Theorem} and~\ref{Universal-Operator-Theorem}, and using the fact that a nonlinearity $\sigma$ is applied following every Kernel Integral layer, it is clear that there exists some learnable $\mathcal{R}_\phi$ which can approximate $\exp{ [\widehat{f}_P \Delta t]}$, and indirectly learnable $v$ which can approximate $\exp{[\widehat{f}_N \Delta t] \Psi_t}$, to an arbitrary margin $\epsilon$. It directly follows that the Fourier Kernel Integral Operator can approximate the first-order SSFM time step update to an arbitrary degree of accuracy which then implies that the FNO as a whole can approximate the solution to the GPE to an arbitrary degree of accuracy, with the number of Kernel Integral layers in the FNO dictating the time discretization $\Delta t$ of the approximation.
\section{Preprocessing photoluminescence datasets}
The datasets are prepared as follows: 
\\
1. Take the absolute value of the original data, PL.
\\
2. A filter is passed to remove the constant white noise with the offsets set to $80\,\mathrm{arb.unit}$. 
\\
3. A logarithmic function is applied to highlight the details, followed by normalization. 
\\
4. Normalized data sets are applied by the contrast-limited adaptive histogram with contrast enhancement limit set to $0.5$, then output data are normalized again. 
\\
5. A cutoff for normalized datasets is applied with offsets $0.5$.
\begin{figure*}[htbp]
\begin{center}
\includegraphics[scale=1.0]{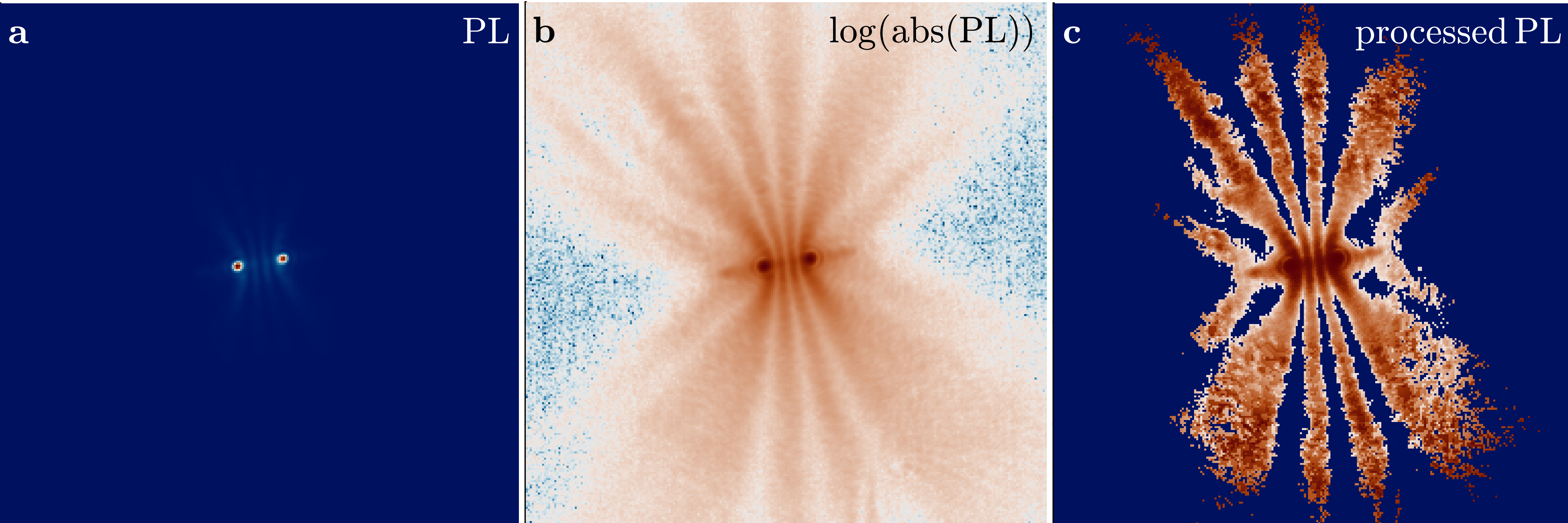}
\caption{\label{comparison_processed_data} \textbf{Comparison of different preprocessed photoluminescence (PL) data.} \textbf{a} Origial PL data. \textbf{b} Logarithmic scale of PL. \textbf{c} Final processed PL.}
\end{center}
\end{figure*}

Figures~\ref{comparison_processed_data}\textbf{a}, \textbf{b} and~\textbf{c} refers to the original PL data, the common treatment shown in the experimental results by applying the logarithmic function, and the one we used here for training the FNO model, respectively.
\section{Hyperparameters}
\begin{figure}[htbp]
\begin{center}
\includegraphics[scale=0.8]{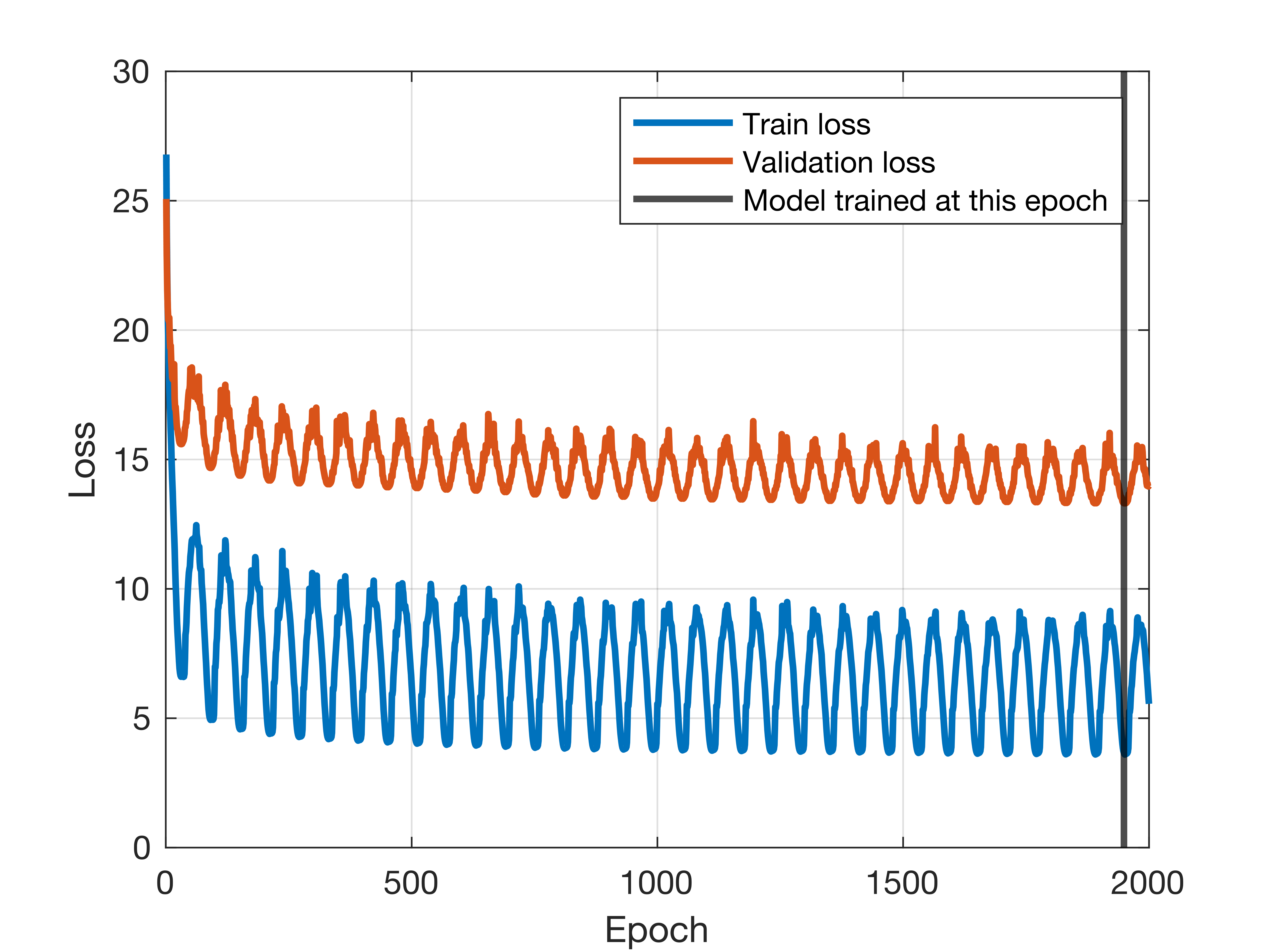}
\caption{\label{S_loss_epoch_theory}\textbf{Training and validation loss vs. epoch for FNO model with theoretically produced datasets.} The black line, indicating the lowest validation loss, is chosen for the purpose of prediction analysis.}
\end{center}
\end{figure}

\begin{figure}[htbp]
\begin{center}
\includegraphics[scale=0.8]{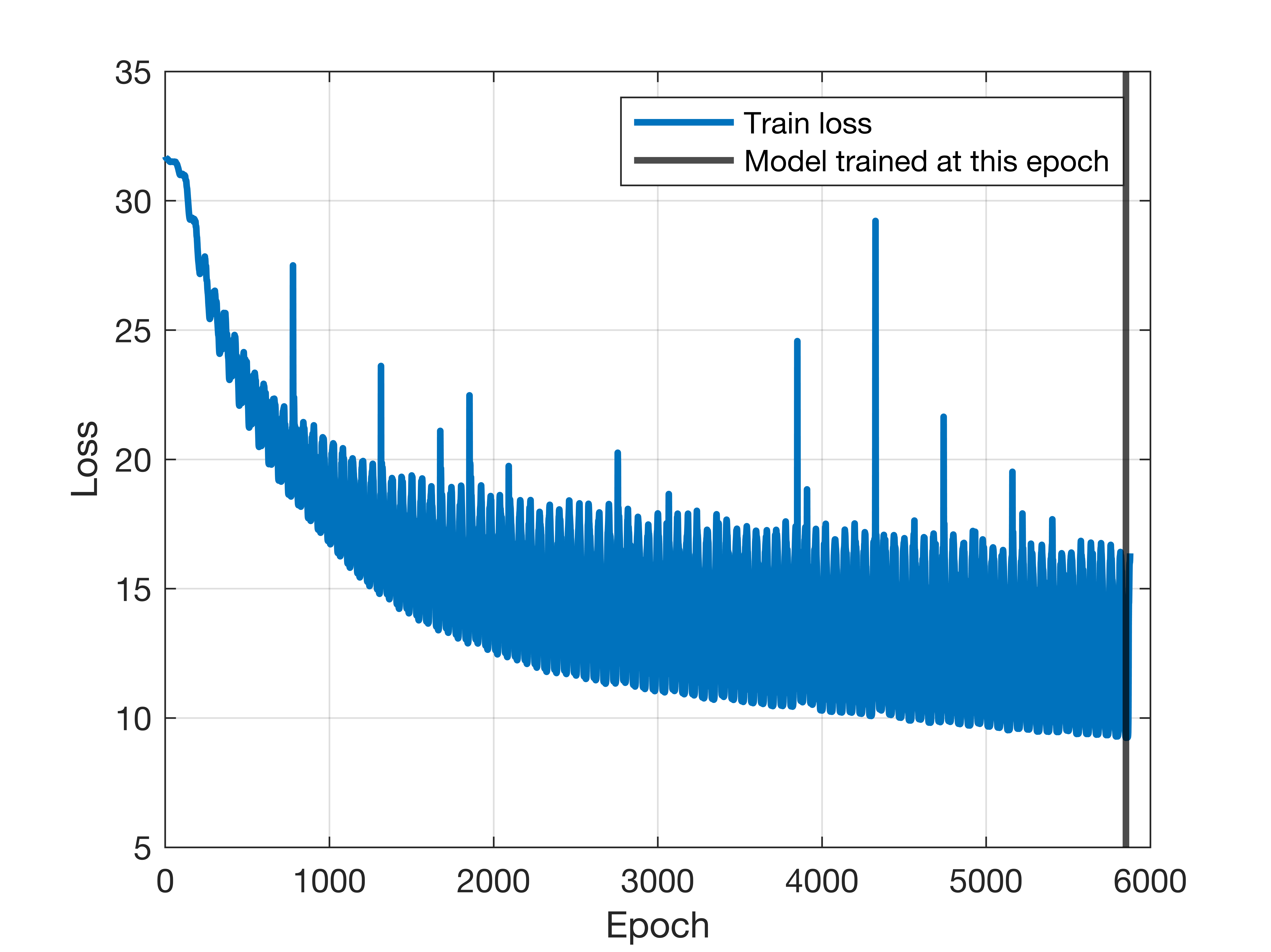}
\caption{\label{S_loss_epoch_experiment}\textbf{Training and validation loss vs. epoch for FNO model with preprocessed experimental datasets.} The black line, indicating the lowest train loss, is chosen for the purpose of prediction analysis.}
\end{center}
\end{figure}

The main hyperparameters for training the FNO model with numerically produced steady-state solutions and experimental observations are shown in Table~\ref{tab:model-hyperparameters-steady-state} and Table~\ref{tab:model-hyperparameters-experiment}, respectively. The loss of training and validation of theoretically produced datasets as a function of epoch is shown in Fig.~\ref{S_loss_epoch_theory}. For experimental results, only the training loss is used shown in Fig.~\ref{S_loss_epoch_experiment}. The numerical datasets, training and testing the ML model are done on a computer with an NVIDIA RTX 4090 GPU and an Intel i9 13900KF.
\begin{table}[htbp]
  \centering
  \caption{Model hyperparameters for steady-state solutions}
  \label{tab:model-hyperparameters-steady-state}
  \begin{tabularx}{\linewidth}{Xl}
    \toprule
    \textbf{Hyperparameter} & \textbf{Value} \\
    \midrule
    Learning Rate & $3\times10^{-3}$ \\
    Weight Decay & $5\times10^{-5}$ \\
    Batch Size & $32$ \\
    Optimizer & Adam \\
    Scheduler & CosineAnnealingLR \\
    Loss Function & H1Loss \\
    Fourier Layers & $4$ \\
    Height Modes & $128$ \\
    Width Modes & $128$ \\
    Hidden Dimension & $64$ \\
    Domain Padding & $0.125$ \\ Domain Padding Mode & symmetric \\
    Activation Function & GeLU \\    \bottomrule
  \end{tabularx}
\end{table}

\begin{table}[htbp]
  \centering
  \caption{Model hyperparameters for experimental datasets}
  \label{tab:model-hyperparameters-experiment}
  \begin{tabularx}{\linewidth}{Xl}
    \toprule
    \textbf{Hyperparameter} & \textbf{Value} \\
    \midrule
    Learning Rate & $3\times10^{-3}$ \\
    Weight Decay & $5\times10^{-5}$ \\
    Batch Size & $32$ \\
    Optimizer & Adam \\
    Scheduler & CosineAnnealingLR \\
    Loss Function & H1Loss \\
    Fourier Layers & $4$ \\
    Height Modes & $128$ \\
    Width Modes & $128$ \\
    Hidden Dimension & $64$ \\
    Domain Padding & $0.125$ \\ Domain Padding Mode & symmetric \\
    Activation Function & GeLU \\    \bottomrule
  \end{tabularx}
\end{table}

\section{Other results for prediction of steady-state solutions}
\begin{figure*}[t]
\begin{center}
\includegraphics[scale=0.85]{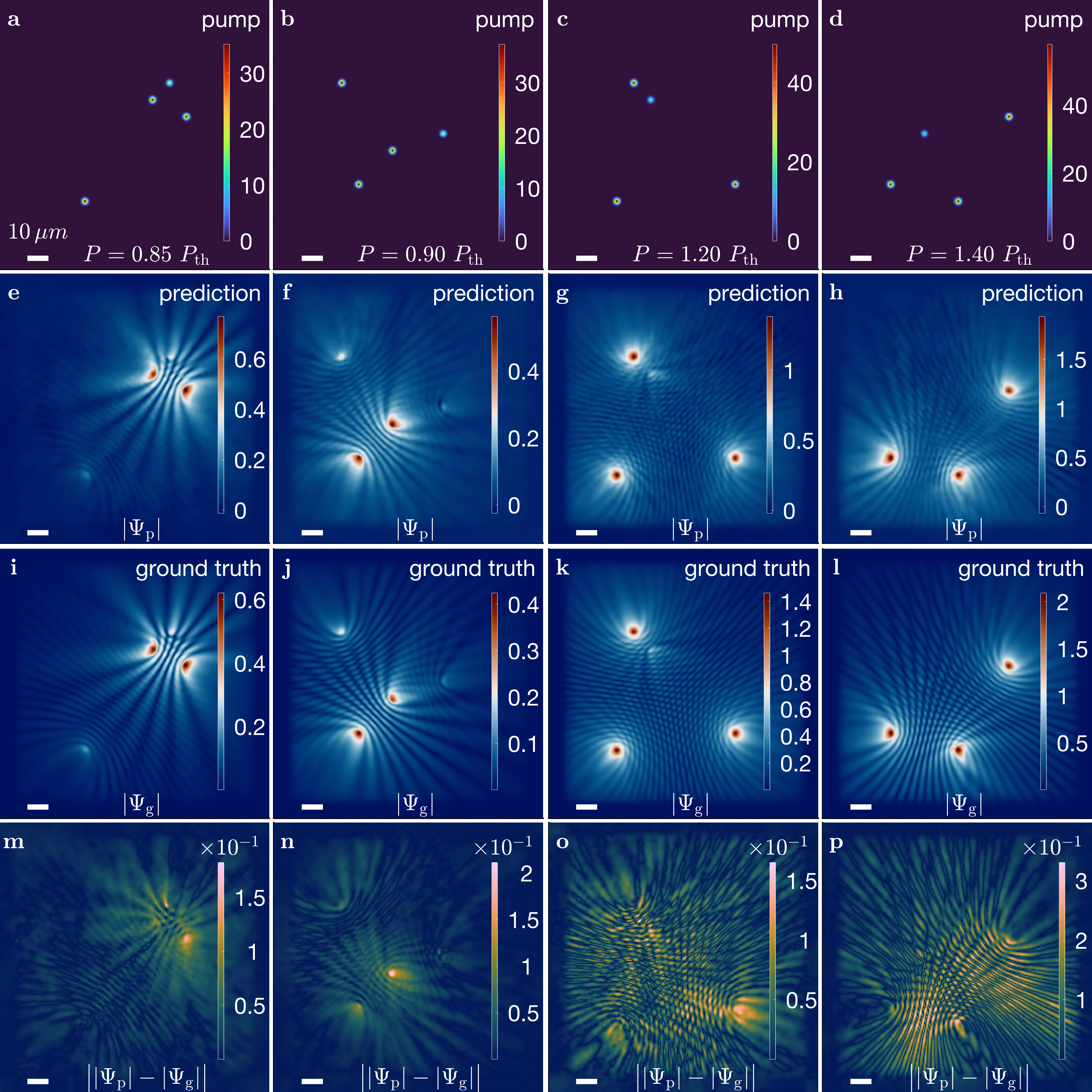}
\caption{\label{S_prediction_fno} \textbf{Comparison of the prediction with large-distanced-spots pump configurations using theoretical datasets and the Fourier Neural Operator approach.} \textbf{a}-\textbf{d} From left to right, the different pump configurations are $P=0.85, 0.9, 1.2, 1.4\,P_{\mathrm{th}}$. \textbf{e}-\textbf{h} Corresponding condensate solutions $|\Psi_p|$ with pump profiles, each of them being different spatial profiles and intensities from the prediction datasets. \textbf{i}-\textbf{l} Corresponding numerical steady-state solutions $|\Psi_g|$ from the ground truth. \textbf{m}-\textbf{p} Corresponding absolute errors between prediction and ground truth $\big||\Psi_p|-|\Psi_g|\big|$. The white bar on all panels is $10\,\mathrm{\mu m}$. The corresponding percentage errors of the condensate particles, taken from Fig.~\ref{S_curve}, are $11.14\%$, $18.07\%$, $3.38\%$, $3.57\%$.}
\end{center}
\end{figure*}

\begin{figure*}[htbp]
\begin{center}
\includegraphics[scale=0.85]{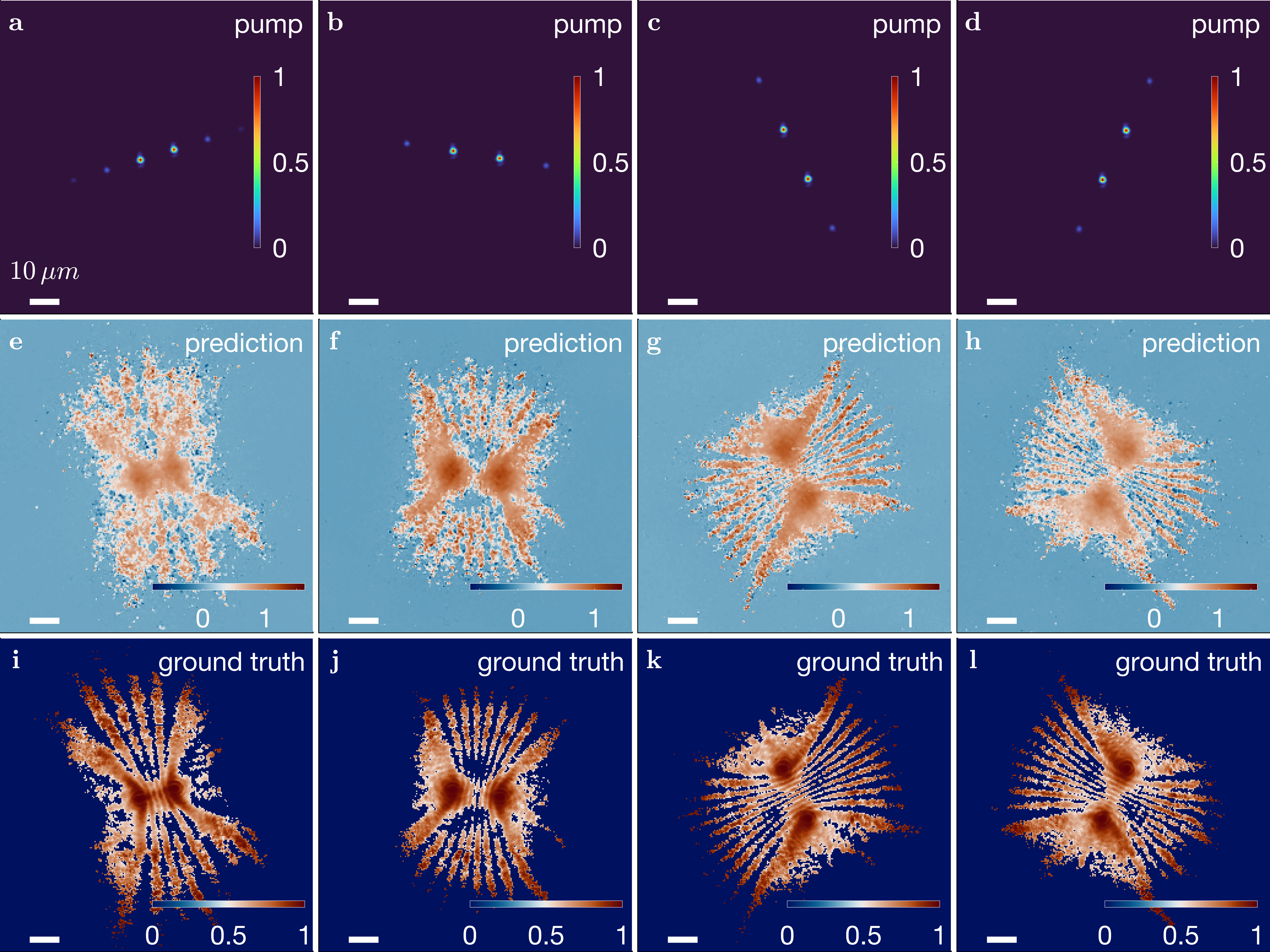}
\caption{\label{S_prediction_fno_experiment} \textbf{Comparison of the prediction with different normalized pump configurations using the preprocessed experimental datasets and Fourier Neural Operator approach.} \textbf{a}-\textbf{d} From left to right, the different pump configurations. \textbf{e}-\textbf{h} Corresponding predictions from the pump profiles.\textbf{i}-\textbf{l} Corresponding post-processed photoluminescence from the experiment. The number of fringes on \textbf{e}-\textbf{h} is $4$, $7$, $11$, $11$, respectively, which is the same as those on \textbf{i}-\textbf{l}. The white bar on all panels is $10\,\mathrm{\mu m}$. The pump density for the experiment is $3.6$ times the threshold of a single spot.}
\end{center}
\end{figure*}
\end{document}